\documentclass[prl,nofootinbib,nobibnotes,twocolumn,showpacs,superscriptaddress]{revtex4}
\usepackage{graphicx,amssymb,bm,latexsym,color,epsf}
\makeatletter
\newcommand{\be}{\begin{equation}}
\newcommand{\ee}{\end{equation}}
\newcommand{\bea}{\begin{eqnarray}}
\newcommand{\ena}{\end{eqnarray}}
\newcommand{\eea}{\end{eqnarray}}

\newcommand{\hs}[1]{\hspace{#1 mm}}
\renewcommand{\a}{\alpha}
\renewcommand{\b}{\beta}
\renewcommand{\c}{\gamma}

\newcommand{\vp}{\varphi}

\newcommand{\nn}{\nonumber\\}

\newcommand{\br}{\bar R}
\newcommand{\bR}{\bar R}
\newcommand{\bg}{\bar g}

\begin{document}

\preprint{KU-TP 064}

\title{A flow equation for $f(R)$ gravity
and some of its exact solutions}
\author{Nobuyoshi Ohta}
\email{ohtan@phys.kindai.ac.jp}
\affiliation{Department of Physics, Kinki University,
Higashi-Osaka, Osaka 577-8502, Japan}
\author{Roberto Percacci}
\email{percacci@sissa.it}
\affiliation{SISSA, via Bonomea 265, I-34136 Trieste, Italy, and INFN, Sezione di Trieste, Italy}
\author{Gian Paolo Vacca}
\email{vacca@bo.infn.it}
\affiliation{INFN, Sezione di Bologna, via Irnerio 46, I-40126 Bologna, Italy}

\begin{abstract}
We write a Renormalization Group (RG) equation for the function $f$
in a theory of gravity in the $f(R)$ truncation.
Our equation differs from previous ones due to the
exponential parametrization of the quantum fluctuations
and to the choice of gauge.
The cutoff procedure depends on three free parameters,
and we find that there exist discrete special choices of parameters
for which the flow equation has fixed points where
$f=f_0+f_1 R+f_2 R^2$.
For other values of the parameters the solution
seems to be continuously deformed.
\end{abstract}

\maketitle

{\it Introduction.}
One way to achieve a quantum-field-theoretic UV completion
of General Relativity is to find a fixed point in a ``theory space''
parametrized by the most general diffeomorphism-invariant semilocal
functional of the metric. The study of the RG flow in this
infinite-dimensional space is a daunting task.
A convenient simplified environment where one can study
the RG flow of gravity is given by effective actions of the form
\be
\Gamma_k=\int d^d x \sqrt{- g} f(R),
\label{action}
\ee
where $f$ depends on the RG scale $k$.
This is called the $f(R)$ truncation.
The first studies focused on the case when $f$ is a polynomial
\cite{lauscherreuter,cpr1,ms,cpr2},
and a fixed point was found there.
This study has now been brought to high order \cite{fallslitim}.
A RG equation for $f$ had been written early on, but serious
attempts to find functional fixed points
(henceforth called ``scaling solutions'')
were not made until \cite{benedetti}.
Negative results concerning the equations written in
\cite{ms,cpr2} have been reported in \cite{dm1,dm2}.
It has been argued that the solution may require truncations
that go beyond functionals of the background field alone,
and taking into account the Ward identities of the
quantum-background split symmetry.
This is a general issue that goes beyond the $f(R)$ truncations
and progress in this direction has been made in
\cite{beckerreuter,dm3}.
In the meantime solutions were found in simplified
(lower-dimensional and/or conformally reduced) settings
\cite{dsz1,dsz2} and recently also in the full four-dimensional case
\cite{dsz3}.

Here we use in the $f(R)$ truncation the lessons learned
in scalar-tensor theories with arbitrary scalar potential
$V(\phi)$ and nonminimal interactions $F(\phi)R$.
It was found there that the use of the exponential parametrization
for the metric and of a ``physical gauge'', based on removing the
spin-zero and spin-one gauge degrees of freedom of the
quantum fluctuation, simplifies the equations to the point where
finding exact scaling solutions in closed form becomes possible
\cite{vacca,LPV}.
While we do not expect the classical equivalence between
$f(R)$ and scalar-tensor theories to extend to the quantum domain,
these results encourage us to try a similar strategy also
in the case of the $f(R)$ truncation.
\bigskip

{\it Flow equation for $f(R)$  Gravity.}
We use the background field method. Motivated by the
nonlinear nature of the space of metrics we split
\cite{nink}:
\bea
g_{\mu\nu}= \bg_{\mu\rho} ( e^h )^\rho{}_\nu\ ,
\label{nonlinear}
\ena
where $\bar g$ is the background field and $h$ the quantum field.
Henceforth we assume that the background space is a sphere.
The York decomposition
$
h_{\mu\nu} = h^{TT}_{\mu\nu} + \nabla_\mu\xi_\nu
+ \nabla_\nu\xi_\mu +
\nabla_\mu \nabla_\nu \sigma
-\frac{1}{d} \bg_{\mu\nu} \nabla^2 \sigma
+
\frac{1}{d} \bg_{\mu\nu} h,
\label{york}
$
where
$\nabla_\mu h^{TT}_{\mu\nu} = \bg^{\mu\nu} h^{TT}_{\mu\nu}
= \nabla_\mu \xi^\mu=0$,
leads to the following very neat form for the off-shell Hessian:
\begin{widetext}
\bea
\label{hessian3}
&&
\frac{1}{2}\int d^d x\sqrt{\bg}\Biggl[
-\frac{1}{2}f'(\bR)h^{TT}_{\mu\nu}\left(-\bar\nabla^2+\frac{2\bR}{d(d-1)}\right)h^{TT\mu\nu}
+s \Bigg(\frac{(d-1)^2}{d^2}f''(\bR)\left(-\bar\nabla^2-\frac{\bR}{d-1}\right)
\nonumber\\
&&
\qquad\qquad\qquad
+\frac{(d-1)(d-2)}{2d^2}f'(\bR)\Bigg)
\left( -\bar\nabla^2-\frac{\bR}{d-1}\right) s
+h\left( \frac{1}{4}f(\bR)-\frac{1}{2d}\bR f'(\bR)\right)h
\Bigg]\ .
\eea
\end{widetext}
Here $s=h-\bar\nabla^2\sigma$ is the gauge-invariant combination of the
spin-zero degrees of freedom.
The only gauge-non-invariant degree of freedom
appearing in the Hessian is the trace part $h$,
and it multiplies the classical equations of motion.

Since we are interested in the off-shell beta functions,
this structure suggests removing $h$ as a gauge choice.
This then leaves a structure that is the same on- and off-shell.
Actually, if we work on the sphere, there is not enough
gauge freedom to remove the constant mode of $h$,
which remains as a single, isolated quantum mechanical mode.
One important virtue of the exponential parametrization is that
results are independent of the way
in which one fixes the residual volume-preserving diffeomorphisms
(with a reparametrization of the trace, this is true
for the full diffeomorphism group \cite{falls}).
We will discuss this point elsewhere.
Here, as in \cite{vacca}, we simply choose the physical
gauge $\xi_\mu=0$.
This leads to a real transverse vector and scalar ghost,
with Lagrangian
$$
C_\mu\left(-\bar\nabla^2-\frac{\bR}{d}\right)C^\mu
+C\left(-\bar\nabla^2-\frac{\bR}{d-1}\right)C\ .
$$
The latter cancels the contribution of the
same operator appearing in the $s$-$s$ Hessian.

To write the flow equation we employ for the spin-$i$ field
a cutoff of the form $f(R)R_k(\Delta_{i})$,
where $R_k(z)=(k^2-z)\theta(k^2-z)$
and $\Delta_0=-\bar\nabla^2-\beta\bar R$,
$\Delta_1=-\bar\nabla^2-\gamma\bar R$,
$\Delta_2=-\bar\nabla^2-\alpha\bar R$.
The parameters $\alpha$, $\beta$, $\gamma$,
introduced in \cite{dsz1}, correspond to different
choices of cutoffs.
According to standard lore, if there is a scaling solution,
its coordinates will generally depend
on such choices but overall qualitative features should not.
By the standard procedure, we get the functional renormalization
group equation
\vskip -0.5cm
\begin{widetext}
\bea
\dot \Gamma_k \hs{-2}&=&\frac{1}{2}
\mbox{Tr}_{(2)}
\left[\frac{\dot f'(\br) R_k(\Delta_2)
+f'(\br) \dot R_k(\Delta_2)}{f'(\br)
\left(P_k(\Delta_2)+\left(\alpha
+\frac{2}{d(d-1)}\right)\br \right)}\right]
-\frac{1}{2}\mbox{Tr}_{(1)}'\left[
\frac{\dot R_k(\Delta_1)}{P_k(\Delta_1)+\left(\gamma-\frac{1}{d}\right)\br} \right]
\nn &&
+\frac{1}{2}\mbox{Tr}_{(0)}'' \left[ \frac{\dot f''(\br)R_k(\Delta_0)
+f''(\br) \dot R_k(\Delta_0)}
{f''(\br) \left(P_k(\Delta_0)+\left(\beta-\frac{1}{d-1}\right)\br \right)+\frac{d-2}{2(d-1)}f'(\br)} \right]\ ,
\label{frge}
\ena
\end{widetext}
where the dot denotes $k\frac{d}{dk}$
and $P_k(z)=z+R_k(z)$.
Here we have not coarse-grained the isolated mode of $h$.
We will comment on this below.
The traces have to be performed summing over the spectra of the
operators $\Delta_i$ using the explicit expression for the eigenvalues $\lambda_n^{(i)}$ and multiplicities $D_n^{(i)}$.
The primes on the traces indicate
that the lowest eigenvalues of $\Delta_1$ and the two lowest eigenvalues of $\Delta_0$ are left out, so that all sums
over the spectral values start from $n=2$ \cite{cpr2}.
The sums can be evaluated using the Euler-Maclaurin formula.
In four dimensions, defining the dimensionless variable $\varphi(r)=k^{-4}f(\bR)$, $r=\bR/k^2$,
one arrives at the following flow equation:
\bea
\!\!\dot\vp&=&\!\! -4\vp+2r \vp'
+\frac{1}{32\pi^2}\Big[\frac{c_1 (\dot\vp'-2r \vp'')+c_2 \vp'}{\vp'[6+(6\a+1)r ]}
\nonumber\\
&&\!\!\!\!\!\!\!\!\!\!\!\!\!\!\!\!\!\!
+ \frac{c_3 (\dot\vp''-2r \vp''')+c_4 \vp''}{\vp''[3+(3\b-1)r]+\vp'}
- \frac{c_5}{4+(4\c-1) r}\Big],
\label{erge}
\eea
where
\bea
&& c_1 = 5+5\Big(3\a-\frac{1}{2}\Big)r+\Big(15\a^2-5\a-\frac{1}{72}\Big) r^2
\nn
&&
+\Big(5\a^3-\frac{5}{2}\a^2-\frac{\a}{72}+\frac{311}{9072}\Big) r^3, \nn
&& c_2= 40+15(6\a-1)r+\Big(60\a^2-20\a-\frac{1}{18}\Big) r^2
\nn&&
+\Big(10\a^3 - 5\a^2-\frac{\a}{36}+\frac{311}{4536}\Big) r^3,
\nn
&& c_3 = \frac{1}{2}\Big[1+\Big(3\b +\frac12 \Big) r+\Big(3\b^2+\b-\frac{511}{360}\Big)r^2
\nn&&
\qquad+\Big(\b^3+\frac{1}{2}\b^2-\frac{511}{360}\b+\frac{3817}{9072}\Big) r^3\Big], \nn
&& c_4 =3+(6\b+1)r+\Big(3\b^2+\b-\frac{511}{360}\Big) r^2, \nn
&& c_5 =12+2(12\c+1) r+\Big(12\c^2+2\c-\frac{607}{180}\Big) r^2.
\nonumber
\ena
Coarse-graining also the isolated mode $h=const$
with the cutoff $k^d$ would produce an additional contribution
$\frac{1}{12\pi^2} \frac{r^2}{16+2\varphi-r\varphi'}$.

The same equation can be derived using the heat kernel expansion.
(Due to the form of the cutoff, only the
coefficients up to $b_6$ are needed.)
We note that the equation is simpler than the ones
that had been written previously.
In particular, compared to \cite{dsz3},
its denominators are of lower order and the function $\varphi$
does not appear undifferentiated in the r.h.s..
In the Einstein-Hilbert truncation, this reduces to the statement
that the cosmological constant does not appear in the flow
of Newton's constant \cite{vacca},
a fact that allows nonsingular flows in the infrared.

\bigskip

{\it Scaling solutions.}
The normal form of the flow equation has a singularity at $r=0$
and further fixed singularities depending on $\beta$.
The analysis of \cite{dm1} showed that isolated solutions
are expected to occur when the number of fixed singularities
matches the order of the equation.
This is the case when $\beta< 0.3945$
(counting only singularities for positive $r$).
Instead of analyzing numerically the equations
for fixed $\alpha$, $\beta$, $\gamma$,
we treat these parameters as unknowns to solve for.
For large $r$ the scaling solutions are expected to grow like $r^2$,
possibly with logarithmic corrections.
The simplest possible solutions are therefore of the form
\be
\varphi(r)=g_0+g_1 r+g_2 r^2\ .
\ee
If we insert this ansatz in the flow equation and write the fixed point equation as a single fraction, its numerator is a fifth order
polynomial in $r$.
By equating to zero the coefficients of all powers of $r$ one
obtains a system of six equations for the six unknowns
$g_0$, $g_1$, $g_2$, $\alpha$, $\beta$ and $\gamma$.
This system has a number of solutions whose properties are
reported in the table.
\medskip

\begin{table}[h]
\begin{tabular}{|r|r|r|r|r|r|l|}
\hline
 $10^3\alpha$&$10^3\beta$ & $10^3\gamma$ &$10^3\tilde g_{0*}$ & $10^3\tilde g_{1*}$ & $10^3\tilde g_{2*}$ & $\theta$ \\
\hline
$-593$ & $-73.5$ & $-177$ & 7.28& $-8.42$ & 1.71 & 3.78 \\
$-616$ & $-70.7$ & $-154$ & 7.42 & $-8.64$ & 1.74 & 3.75  \\
$-564$ & $-80.3$ & $-168$ & 6.82 & $-8.77$ & 1.83 & 3.70 \\
$-543$ & $-87.4$ & $-126$ & 6.31 & $-9.47$ & 2.06 & 3.43 \\
$-420$ & $-100.5$ & $-3.19$ & 4.90 & $-10.2$ & 2.83  & 2.93 \\
$-173$ & $-2.98$ & 244 & 4.53& $-8.34$ & 2.70  & 2.18 \\
\hline
$-146$ & $-64973$ & 250 & 2.90 & $-10.7$ & 0.0006  & 2.58  \\
$-109$ & $-22267$ & 307 & 2.90 & $-10.4$ & 0.0045 & 2.45 \\
\hline
109 & $-3564$ & 526 & 2.84 & $-7.83$ & 0.094 & C  \\
377 & $-1305$ & 794 & 2.57 & $-4.37$ & 0.214 & $>4$   \\
\hline
\end{tabular}
\caption{The properties of the exact quadratic solutions: parameters (first three columns), couplings (nest three columns) and critical exponent (last column). "C" stands for "complex.}
\end{table}

\medskip
Since the denominator of the equation has at least one zero for positive $r$
that is not matched by a corresponding zero of the numerator,
the solutions shown are valid on the whole positive real axis
except for isolated points.

All fixed points have a critical exponent that is exactly
equal to 4, corresponding to the volume degree of freedom.
The last column reports the most relevant exponent
besides this, obtained from an analysis based on a
polynomial expansion around the origin.
The first six lines are solutions with exactly two relevant directions; the two subsequent ones have three relevant directions.
The last two lines have four relevant directions,
of which two are complex conjugate (denoted by C); furthermore
the last has a critical exponent that is larger than four.
The other solutions have only real critical exponents,
contrary to previous analyses in full $f(R$) gravity.
We consider this a desirable feature.

We observe that for all these scaling solutions,
the equation of motion $2f-Rf'=0$ has the solution
$R_*=-2 g_0 k^2/g_1$, and therefore avoid the possible issue
of the redundancy of all eigenperturbations \cite{dm2}.
Since $g_0>0$ and $g_1<0$
the solution is compatible with spherical topology.

The numerical similarities between the six solutions of the first group
and the two solutions of the second group suggests that perhaps they
are the same scaling solution in different cutoff schemes.
(The quantitatively non-negligible differences by factors
of order one are typical for the
scheme dependence in this type of calculations.)
To substantiate this hypothesis we have studied the polynomial
approximation to the flow equation along straight paths in
the space of parameters joining the different solutions.
(For example, $\alpha(t)=-0.593 - 0.0227 t$,
$\beta(t)=-0.0735 + 0.00281 t$,$
\gamma(t)=-0.177 + 0.0226 t$
interpolates between the first two solutions when $0\leq t\leq1$.)
We have considered truncations to sixth order polynomials,
and find that for any pair of solutions in the first group
there exists a continuous interpolation.
Likewise, the two solutions of the second group are continuously
related.

In addition to the expansion around $r=0$ we have also considered
the asymptotic behavior of the solutions for large $r$.
For generic values of $\alpha$, $\beta$, $\gamma$ it is of the form
$$
\varphi(r) = A r^2+ a_1 r +a_0 +a_{-1}/r +a_{-2}/r^2 +\ldots
$$
where $A$ is a free parameter and $a_i$ are
given functions of $\alpha$, $\beta$, $\gamma$.
For $\alpha=1/6$ also the parameter $A$ is fixed.
For $\beta=0$ or $\gamma=1/4$ the leading term goes like
$r^2 \log{r}$ and also subsequent terms have logarithmic corrections.
It will clearly be important to match these behaviors and
establish numerically the existence
of scaling solutions of (\ref{frge}) for other cutoff choices.

While we do not have an exact, nor a complete numerical
solution except at the special points listed,
the existence of continuous polynomial
deformations suggests very strongly that
the existence of the scaling solutions is not limited
to the special cutoff choices listed in the table.
Instead, we conjecture that the scaling solution
exists in a whole open subspace of parameters
and that the effect of choosing special values of
the parameters is to give the scaling solution
a very simple functional form.

We have also looked for polynomial solutions with
more traditional cutoff choices.
With the so-called type I cutoff
$\alpha=\beta=\gamma=0$,
truncating at order eight polynomial,
we find the solution
\bea
\varphi(r)&=&0.00482
-0.00463r
+0.00131r^2
\nn
&&
\!\!\!\!\!\!\!\!\!\!\!
\!\!\!\!\!\!\!\!\!\!\!
-1.86\times10^{-5}r^3
-5.02\times10^{-6}r^4
-9.11\times10^{-7}r^5
\nn
&&
\!\!\!\!\!\!\!\!\!\!\!
\!\!\!\!\!\!\!\!\!\!\!
-2.00\times10^{-7}r^6
-4.23\times10^{-8}r^7
-9.51\times10^{-9}r^8
\nonumber
\eea
with critical exponents
$4$, $2.21$, $-2.51$, $-5.21$, $-7.50$, $-9.52$,
$-11.45$, $-12.9$, $-15.1$.
This can be seen to be a continuous deformation of the first group of
fixed points in the table.

In \cite{cpr2} a type-I cutoff was also used, so it is useful
to compare this solution to the one given in tables III and IV there.
The differences, which can be ascribed to the different parametrization
of the quantum field and to the different gauge choice,
are significant.
While $g_0$ is almost the same, $g_1$ is here smaller by a factor five,
which means that $\tilde G_*$ is larger by a comparable factor.
This is in accordance with the results of \cite{vacca}.
Whereas in \cite{cpr2} the coefficients $g_i$
remained of comparable size
as functions of $i$, here they decrease quite strongly.
This is similar to the findings of \cite{eichhorn}
in unimodular gravity.
Most significantly, there are here only two relevant directions
instead of three, as found so far in all treatments
of the $f(R)$ truncation.

In the same truncation but with the type-II cutoff
$\alpha=-1/6$, $\beta=1/3$, $\gamma=1/4$,
deforming the first group of solutions
leads to the fixed point
\bea
\varphi(r)&=&0.00426
-0.00702r
+0.00280r^2
\nn
&&
\!\!\!\!\!\!\!\!\!\!\!
\!\!\!\!\!\!\!\!\!\!\!
-2.60\times10^{-4}r^3
-1.31\times10^{-5}r^4
-7.06\times10^{-7}r^5
\nn
&&
\!\!\!\!\!\!\!\!\!\!\!
\!\!\!\!\!\!\!\!\!\!\!
+5.47\times10^{-8}r^6
+1.43\times10^{-9}r^7
+8.31\times10^{-10}r^8
\nonumber
\eea
with critical exponents
$4$, $1.91$, $-1.54$, $-4.73$, $-7.52$, $-10.0$,
$-12.4$, $-14.3$, $-23.3$.

From a preliminary analysis of polynomial expansions,
it would seem that the solutions of the second group
cannot be continued to type I or type II cutoffs.

\bigskip

{\it Alternative equation.}
Performing the spectral sums as in \cite{benedetti},
leads again to equation (\ref{frge})
but with different coefficients:
\bea
&& c_1 = \frac{5}{108}(6+(6\alpha+1)r)\Big(18+3(12\alpha-5)r
\nn&&
+(2-15\alpha+18\alpha^2)r^2\Big)
\nn
&&
c_2= \frac{5}{108} \Big(864+18(108\alpha-17)r
+3(432\alpha^2\nn&&-150\alpha+1)r^2
+2\left(108\alpha^3-72\alpha^2-3\alpha+2\right)r^3\Big)
\nn
&& c_3 = \frac{1}{72}\Big(36+12(9\beta+1)r
+(108\beta^2+24\beta
\nn&&
-53)r^2
+\left(36\beta^3+12\beta^2-53\beta+15\right)
r^3\Big)
\nn
&& c_4 =3+\left(6\beta+\frac{5}{4}\right)r
+\left(3\beta^2+\frac{5}{4}\beta-\frac{11}{8}\right)r^2,
\nn
&&
c_5 =12+3(8\gamma+1)r
+\left(12\gamma^2+3\gamma-\frac{19}{6}\right)r^2
.
\nonumber
\ena
This equation admits again some exact quadratic solutions.
The interesting ones are
\medskip

\begin{tabular}{|r|r|r|r|r|r|l|}
\hline
 $10^3\alpha$&$10^3\beta$ & $10^3\gamma$ &$10^3\tilde g_{0*}$ & $10^3\tilde g_{1*}$ & $10^3\tilde g_{2*}$ & $\theta$ \\
\hline
$-97.8$ & 38.9 & 319 & 4.31 & $-7.46$ & 2.85 & $2.03$ \\
\hline
$-438$ & $-122$ & $-21.0$ & 4.67 & $-10.4$ & 3.14  & $3.2$  \\
\hline
134 & $-2765$ & 551 & 2.82 & $-7.70$ & 0.13  &  C \\
505 & $-715$ & 922 & 2.16 & $-2.65$ & 0.21 & $>4$ \\
$-564$ & $-63.8$ & $-147$ & 7.83 & $-6.80$ & 1.35 & $>4$  \\
\hline
\end{tabular}
\medskip

The first has two relevant directions and is the analogue of the
first six solutions of (\ref{erge});
the following one has three relevant directions;
the third has four relevant directions and the last two have at least one critical exponent larger than four,
possibly a complex conjugate pair, but with very slow convergence with the order of the polynomial.

Much of what was said for the other equation can be repeated with small
changes in this case.
\bigskip

{\it Discussion.}
The main positive result of this work is
the existence of special quadratic scaling solutions
for particular cutoff choices, which remain close to this shape
when the cutoff is varied generically.
These solutions provide further evidence for asymptotic safety
of gravity.
The fact that they show similarities
to the Starobinsky model of inflation (which is favored
by current cosmological data) is also encouraging.

While these results definitely constitute progress relative to
polynomial truncations, the situation is still not satisfactory.
The existence of physically undesirable
solutions with critical exponents larger than four
suggest that these may be artifacts of the approximations made
and that they may go away in an improved treatment.
More generally, the existence of multiple solutions
makes the identification of the physically relevant one not obvious.
In this respect there is of course the ever-present need to go
beyond the
single-field truncations and to take into account the
split-symmetry Ward identities.
But even within the single-field truncation there may be room for
better understanding.
One aspect that remains somewhat unclear is the physical meaning
of the limit $r\to\infty$, ($k^2\ll R$)
i.e. of coarse-graining at length scales that are much larger
than the diameter of the manifold.
In fact it had been suggested in \cite{dsz2} that one may not even
need to solve any equation beyond some finite value of $r$.
Here, as in \cite{dsz3}, we do solve an equation for all $0<r<\infty$,
but for both equations we have considered,
their form for large $r$ is somewhat ambiguous.
This is because both the use of the Euler-Maclaurin formula and the
method of \cite{benedetti} are two arbitrary ways of
analytically interpolating the spectral sums.
The existence of isolated singular points in
our solutions may be a manifestation of the same problem.
We observe that these issues could be avoided using a non-compact background manifold.
\bigskip

{\it Acknowledgment.}
This work was supported in part by the Grant-in-Aid for
Scientific Research Fund of the JSPS (C) No. 24540290.

\end{document}